\documentclass[twocolumn, aps, superscriptaddress, groupedaddress]{revtex4}  

\newcommand{\bra}[1]{\langle #1|}  
\newcommand{\ket}[1]{|#1\rangle}

\usepackage{mathtools}
 
\usepackage{amsmath,amsthm,amsfonts,amssymb}   
\usepackage{fontenc}
\usepackage[all]{xy}
\usepackage{graphicx,subfigure,multirow}
\usepackage{tikz}
\usetikzlibrary{shapes, arrows, shadows} 
\usepackage[latin1]{inputenc}
\usepackage[scaled]{helvet}
\usepackage[toc, page]{appendix}
\usepackage{graphicx}				
\usepackage{amssymb}

\newtheorem{result}{Result}

\begin{document} 

\title{Towards device-independent information processing on general quantum networks}   
\author{Ciar{\'a}n~M. Lee} 
\email{ciaran.lee@ucl.ac.uk}
\affiliation{Department of Physics and Astronomy, University College London, Gower Street, London WC1E 6BT, UK} 
\author{Matty~J. Hoban}
\affiliation{University of Oxford, Department of Computer Science, Wolfson Building, Parks Road, OX1 3QD, UK.}
\affiliation{School of Informatics, University of Edinburgh, 10 Crichton Street, Edinburgh EH8 9AB, UK}

\begin{abstract}   
The violation of certain Bell inequalities allows for device-independent information processing secure against non-signalling eavesdroppers. However, this only holds for the Bell network, in which two or more agents perform local measurements on a single shared source of entanglement. To overcome the practical constraint that entangled systems can only be transmitted over relatively short distances, large-scale multi-source networks have been employed. Do there exist analogues of Bell inequalities for such networks, whose violation is a resource for device-independence? 
In this paper, the violation of recently derived polynomial Bell inequalities will be shown to allow for device-independence on multi-source networks, secure against non-signalling eavesdroppers. 

\end{abstract} 

\maketitle   

The violation of Bell inequalities has 
been shown to have immense practical importance for quantum information processing \cite{barrett2005no, brunner2014bell, acin2007device}. Indeed, violation of certain Bell inequalities is a resource for unconditionally secure key distribution \cite{barrett2005no, brunner2014bell, vazirani2014fully, acin2007device, barrett2012unconditionally, acin2006bell, pironio2009device} and randomness amplification \cite{colbeck2012free, ramanathan2016randomness, pironio2010random}, and for achieving certain computational advantages \cite{anders2009computational, hoban2011non, hoban2011generalized, wallman2014nonlocality}\footnote{See also \cite{lee2016deriving} for a connection between other physical phenomena and computation.}. Moreover, these protocols are \emph{device-independent}, meaning they depend only on the observed output statistics of devices used to implement them. 
\color{black}{In the case of key distribution, for certain protocols the violation of a Bell inequality can be used to lower bound the secure key rate \cite{acin2007device}. Furthermore,} \color{black} monogamy relations have been derived between the violation of certain Bell inequalities and the amount information an eavesdropper can obtain about the generated key; the higher the violation, the lower the information \cite{barrett2006maximally, aolita2012fully, colbeck2008hidden}. 
  
\begin{figure}[t] 
\begin{subfigure}
\centering
(a)\includegraphics[scale=0.1]{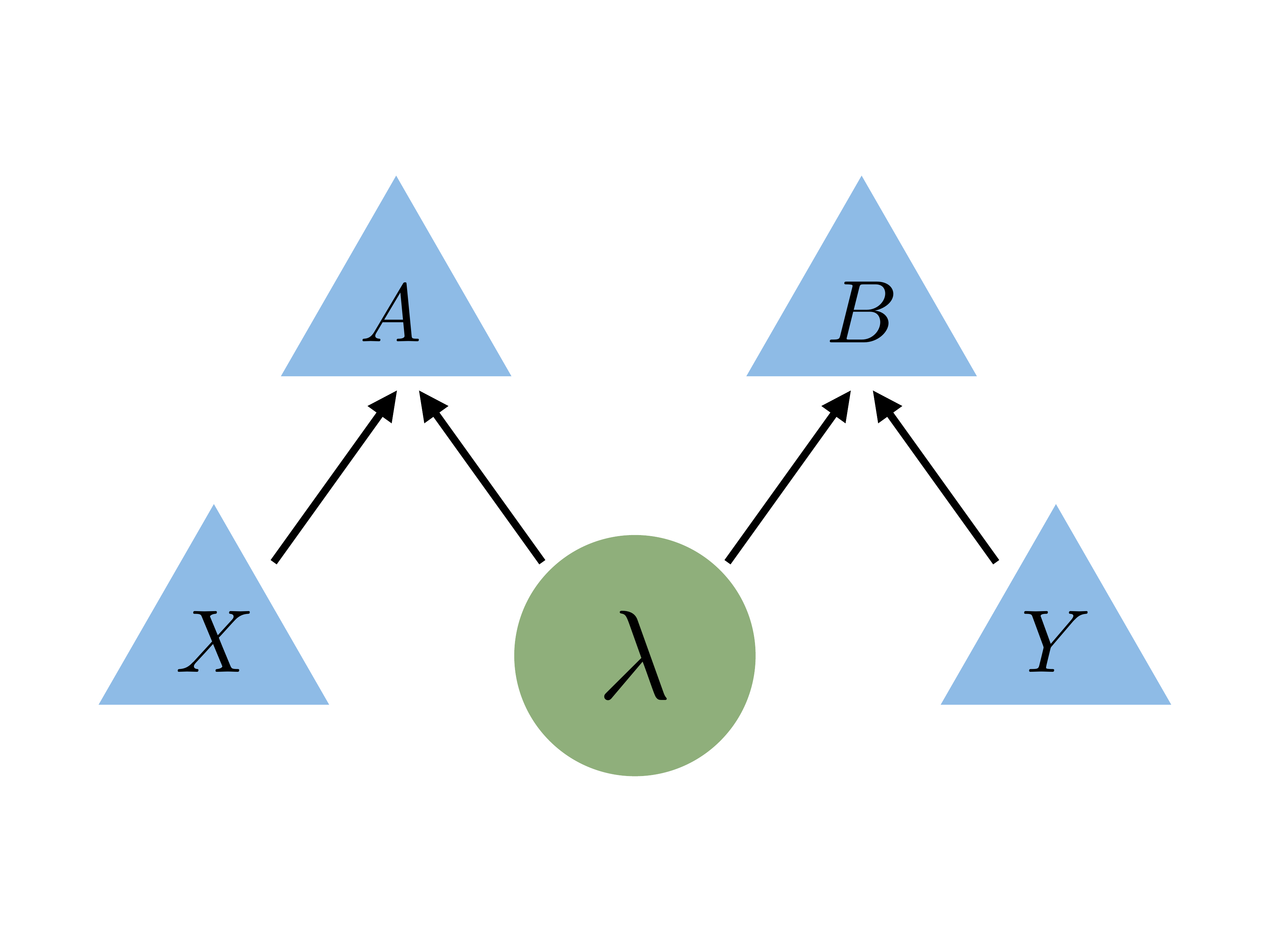}
\end{subfigure}
\begin{subfigure}
\centering
(b)\includegraphics[scale=0.1]{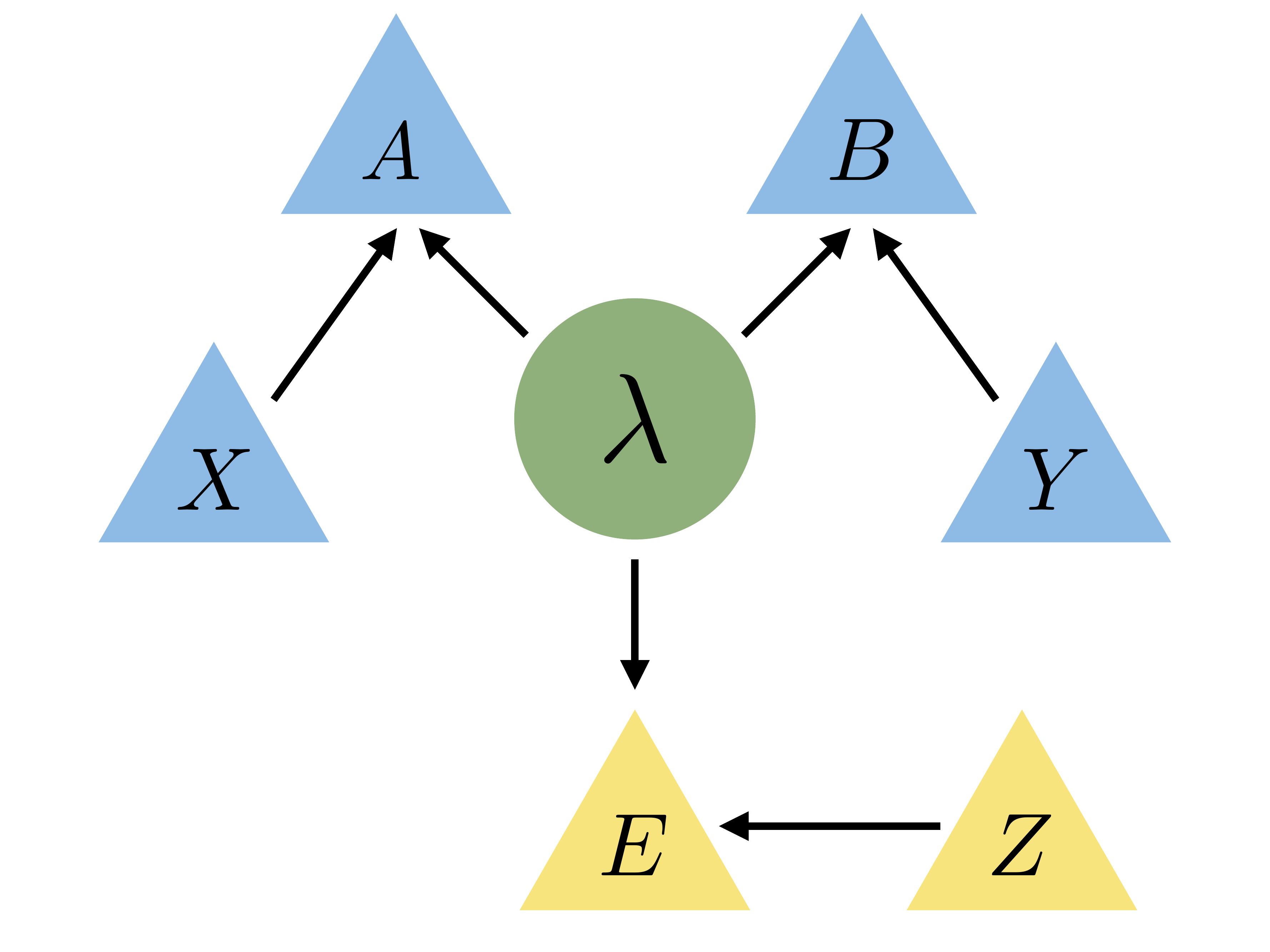}
\end{subfigure}
\caption{
(a) Bell network: $X,Y$ and $A,B$ denote inputs and outcomes of the agents. (b) Bell network with eavesdropper.
}
\label{Bell scenario}
\end{figure}

\begin{figure}[t] 
\begin{subfigure}
\centering
(a)\includegraphics[scale=0.105]{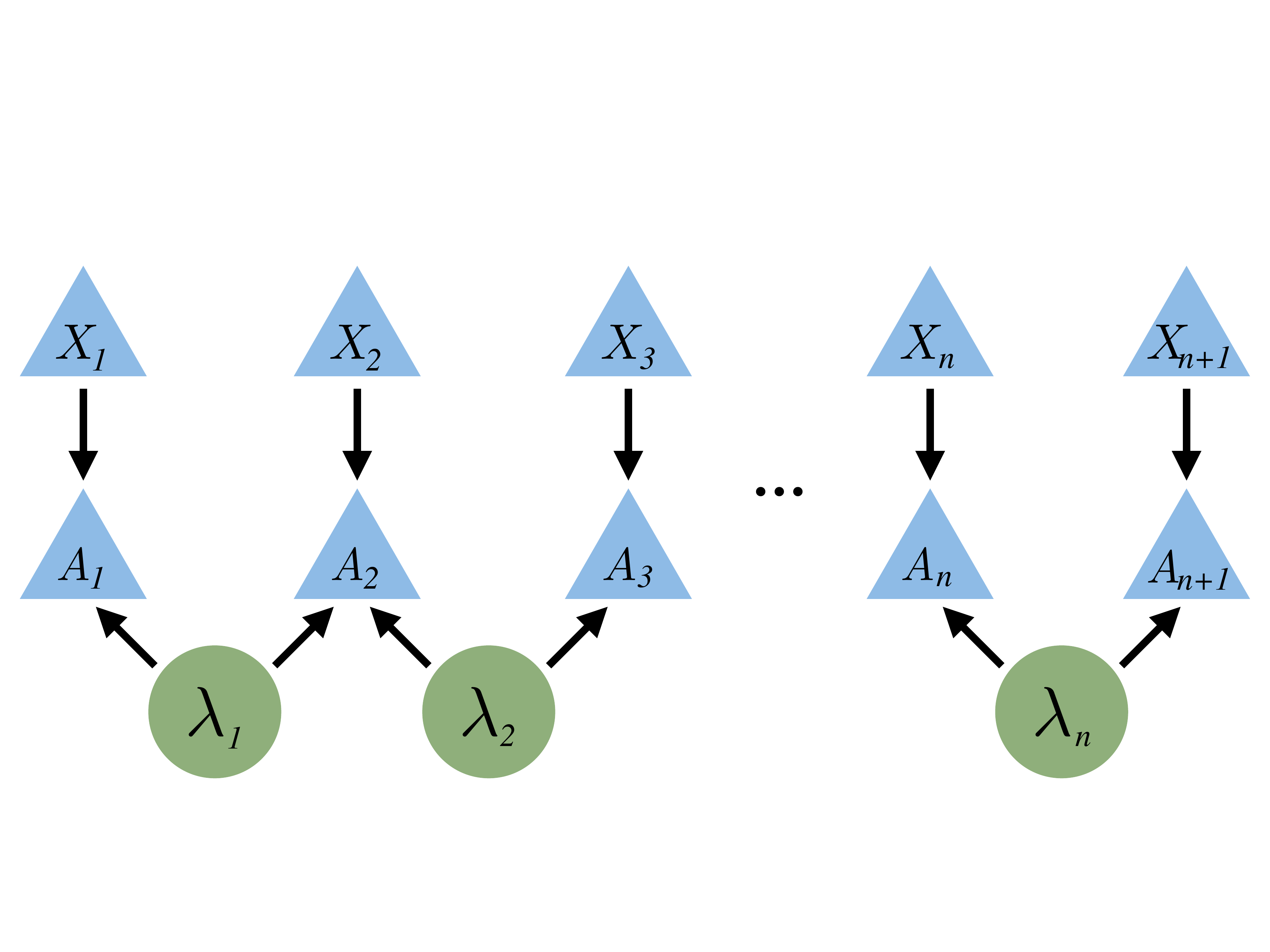}
\end{subfigure}
\begin{subfigure}  
\centering
(b)\includegraphics[scale=0.105]{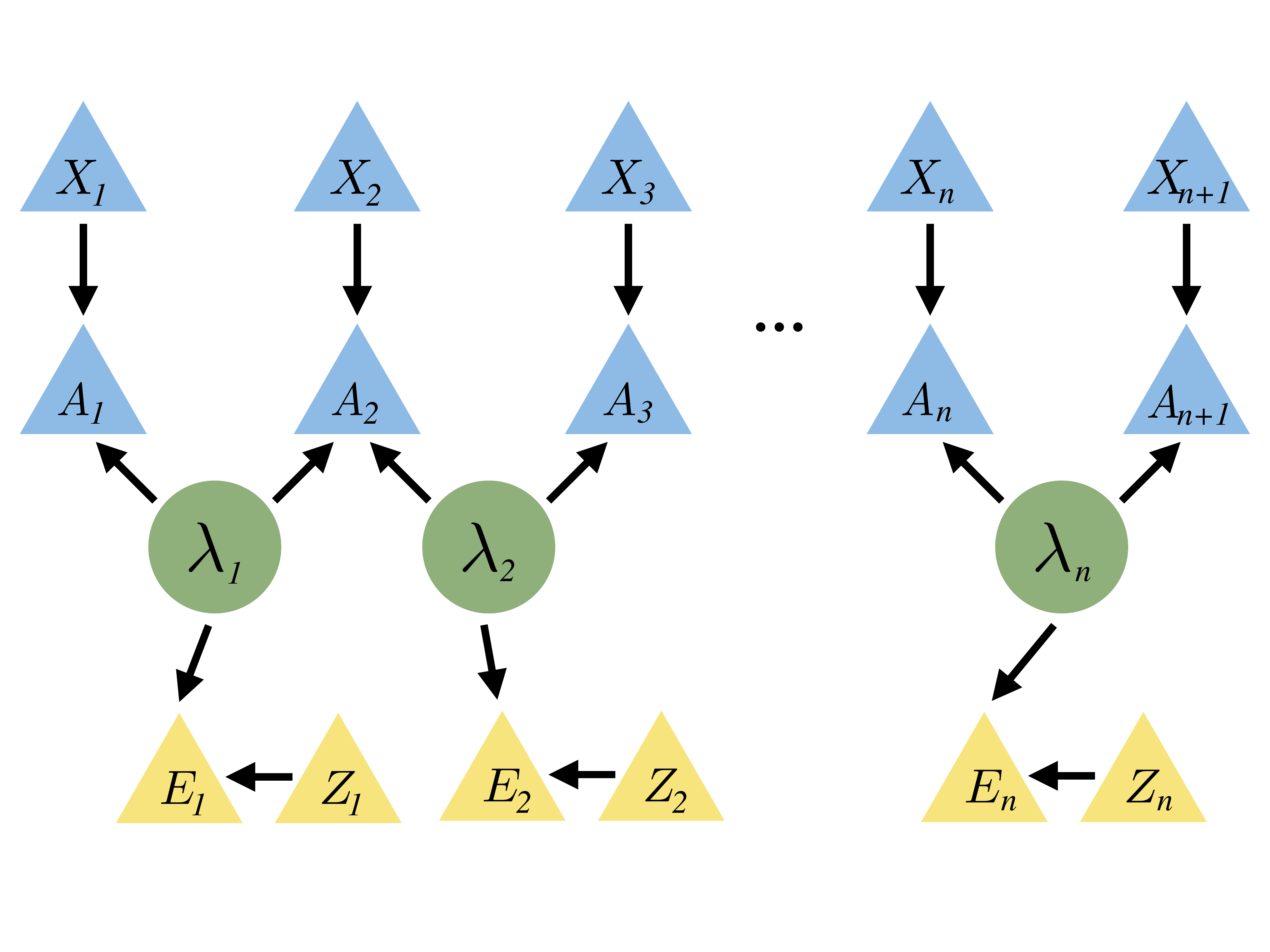}
\end{subfigure}
\begin{subfigure}   
\centering
(c)\includegraphics[scale=0.104]{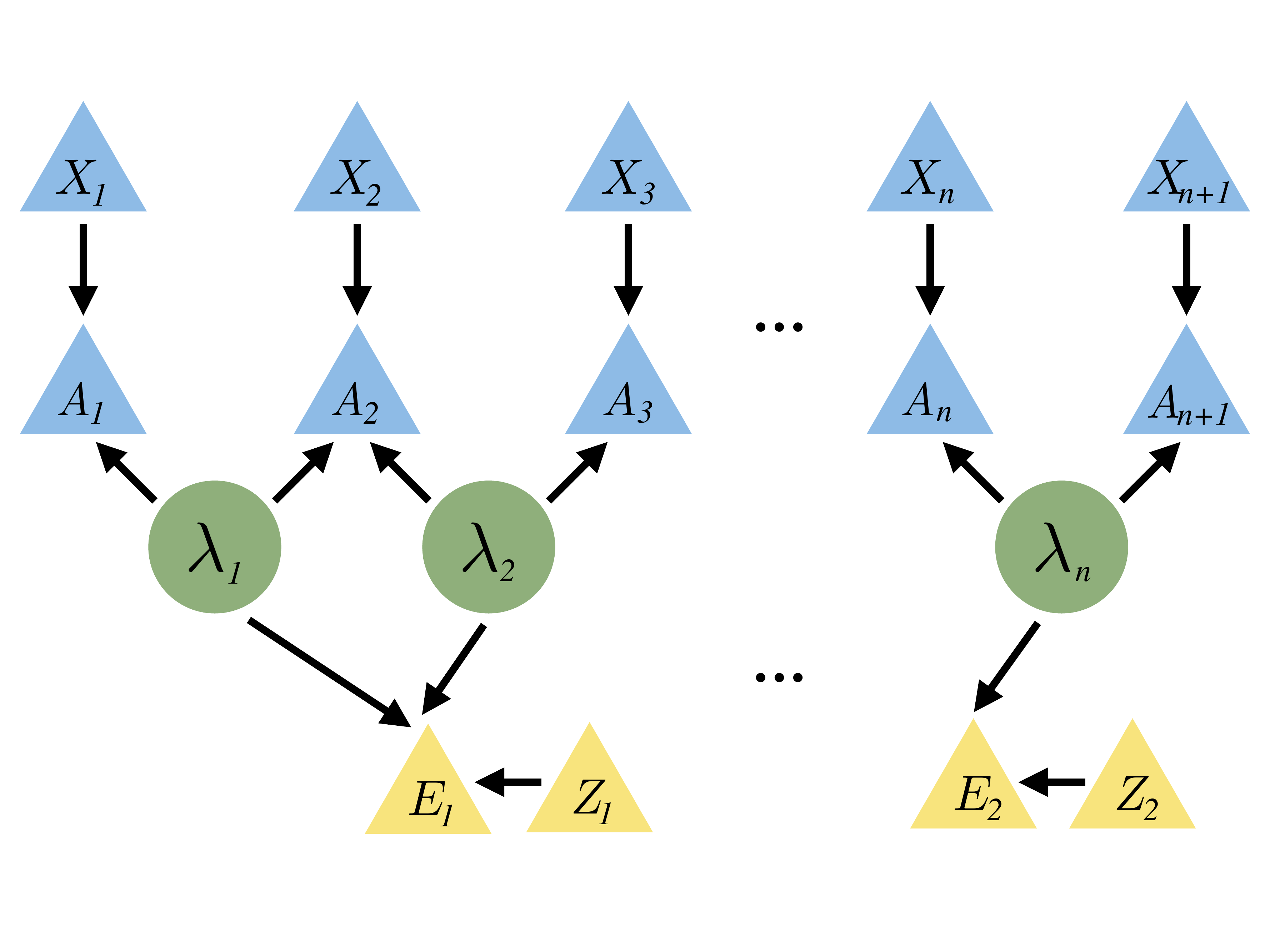}
\end{subfigure} 
\begin{subfigure}
\centering 
(d) \includegraphics[scale=0.1035]{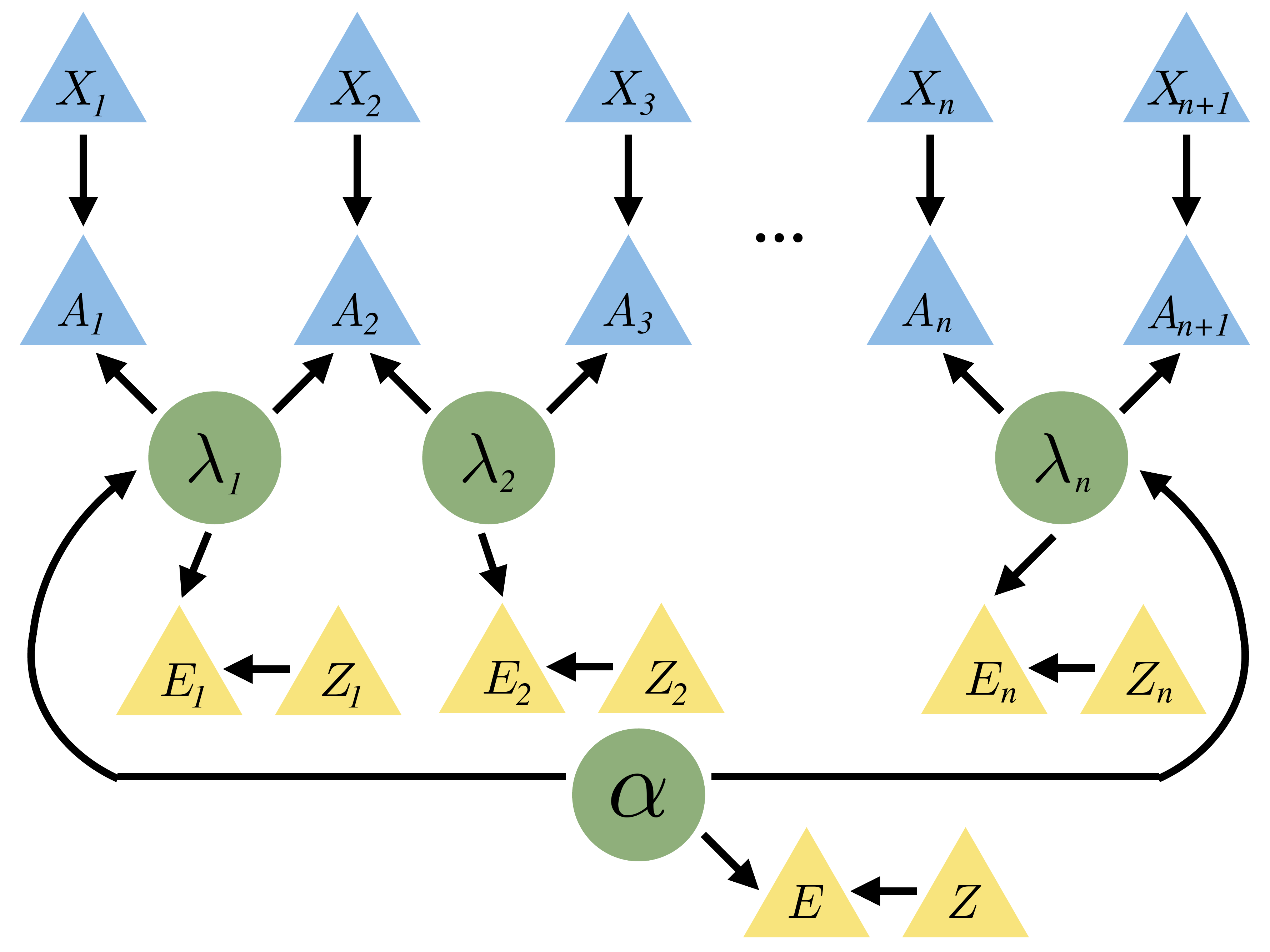}  
\end{subfigure}
\caption{
(a) Repeater network: $x_i$ and $A_i$ denote the possible inputs and outcomes of agent $i$. (b) Repeater network with eavesdropper holding a system correlated with each source: $z_i$ and $E_i$ denote the possible inputs and outcomes of the measurement on each eavesdropper's system. (c) Eavesdropper holding system correlated with all but last source.  (d) Eavesdropper correlating first and last sources. 
}
\label{Repeater network}  
\end{figure}

However, these results only hold for the Bell network depicted in Fig.~\ref{Bell scenario}(a), in which two agents perform local measurements on a single shared source of entangled systems. The utility of these networks is limited by practical constraints: entangled systems can only be transmitted over relatively short distances, and only a small number of agents can share an entangled state distributed by a single source \cite{sangouard2011quantum, seri2017quantum, yin2017satelite}. To overcome this, large-scale multi-source quantum networks, such as that schematically illustrated in Fig.~\ref{Repeater network}, have been employed \cite{sangouard2011quantum, seri2017quantum, van2017multiplexed, reiserer2016robust, goodenough2016assessing, rozpedek2017relaistic, satoh2017network, sun2016quantum, valivarthi2016quantum, pirandola2017fundamental, pirandola2016capacities}. 
Yet having multiple intermediate nodes in the network opens the door for novel eavesdropping attacks not seen in the Bell network. 
Do there exist analogues of Bell inequalities for such multi-source networks whose violation is a resource for device-independence
and which can prevent novel eavesdropper attacks?


Recently, \emph{polynomial Bell inequalities} have been derived \cite{chaves2016polynomial, rosset2016nonlinear, wolfe2016inflation, lee2015causal, branciard2012bilocal, branciard2010characterizing, tavakoli2016bell, tavakoli2014nonlocal, mukherjee2015correlations} on the correlations classically achievable in multi-source networks. \color{black} Violation of these polynomial inequalities witnesses non-classical behaviour in such networks. Can such violations be connected to advantages in information processing on large quantum networks, as was the case in the Bell network? \color{black} 
The main obstacle to establishing such a connection is that the set of classical correlations of a given general network forms a non-convex semi-algebraic set \cite{lee2015causal}, thus methods establishing standard Bell inequality violation as an information-theoretic resource are no longer applicable \footnote{Although, it should be noted that there do exist families of networks more general than the Bell network---involving sequenital measurements on a single shared source of entanglement---that each give rise to a convex set of classical correlations \cite{gallegoseq,curchod2017}.}.   
 
Such bounds on the correlations generated in multi-source networks were originally studied in the field of Causal Inference \cite{Pearl-09, Spirtes-Glymour-01}. Recently, the tools and formalism pioneered in this field have begun to see applications in quantum information \cite{wood2015lesson, allen2016quantum, chaves2015device, chaves2015unifying, chaves2016causal, brask2017bell}. Indeed, this formalism subsumes and generalises standard cryptographic constraints such as no-signalling and the assumption that agents each have a secure laboratory, as is discussed in more detail in Appendix~\ref{DAGs}. In this formalism, agents and sources are represented by nodes in directed acyclic graphs (DAGs), with the arrows denoting the causal relationship between nodes. For example, the DAG in Fig~\ref{Bell scenario}(a) corresponds to the causal structure of the Bell network. \color{black} As in standard device-independence, this work assumes that both measurement devices used by agents, and the sources they measure, are supplied by an untrusted adversary, or eavesdropper \footnote{Furthermore, it is assumed here that an eavesdropper can only apply \textit{individual attacks}, where the eavesdropper's attack on each run of the experiment is independent of the attacks on other runs. Moreover, systems are assumed to be independently and identically generated, and finite statistical effects will not be taken into account. These assumptions allow the study of device-independent cryptography in multi-source networks to be initiated. It is left open for future work to prove security under minimal assumptions.}. 

In the Bell network, violation of a Bell inequality rules out quantum \cite{vazirani2014fully} and post-quantum \cite{barrett2012unconditionally,acin2007device} eavesdroppers. In these attacks, the systems measured by the agents could be correlated with a system held by the eavesdropper, as depicted in Fig.~\ref{Bell scenario}(b). These need not be quantum systems, as long as the DAG of Fig.~\ref{Bell scenario}(b) holds. Measuring this system using a device with input $Z$ and outcome $E$, an eavesdropper could gain information about agents' outcomes. That is, it may be that $P(E|A, B, Z)\neq P(E|Z)$. By violating the chained Bell inequality however \cite{barrett2006maximally, colbeck2008hidden}, agents can limit the information such an eavesdropper can gain about their outcomes.
   
However, for the general networks considered here, there are new avenues for eavesdropping attacks. 
The eavesdropper supplying of the sources can hold systems correlated with more than a single source, as in Fig.~\ref{Repeater network}(c). Additionally, the eavesdropper could even introduce correlations \footnote{Recall that each agent is in a lab isolated from the rest of the world. All they observe is their device outcome. They are completely ignorant about whether the eavesdropper has manufactured the specified independent sources---or is trying to cheat by introducing correlations between sources.} between sources, as in Fig.~\ref{Repeater network}(d). The main findings of the current work are: 
\begin{enumerate} 
\item For an eavesdropper holding independent post-quantum systems that can each be correlated with a single source, the violation of certain polynomial inequalities bounds this eavesdropper's information about agents' outcomes. 
\item By introducing correlations between sources assumed by the agents to be independent, an eavesdropper can simulate quantum correlations consistent with the original DAG, hence gaining complete knowledge of agents' outcomes. However, increasing the measurement settings can combat this.
\item A new intermediate device-independence scenario: trusting a subset of sources are not correlated with a single system held by an eavesdropper. Given this assumption, the attack of item $2$ can be prevented.
\end{enumerate} 
It should be emphasized here that, as in the work of Ref.~\cite{barrett2012unconditionally}, security will be established against \emph{non-signalling} eavesdroppers by bounding their predictive power to learn outcome of agents' devices. In the case of the Fig.~\ref{Bell scenario}(b), letting $D\left(P,Q\right):=\frac{1}{2}\sum_x |P(x)-Q(x)|$ denote the total variational distance, this corresponds to bounding $D(P(E|A, X, Z), P(E|Z))$. Another way of bounding this is through the \textit{device-independent guessing probability} (see e.g. \cite{masanes2011}). That is, through establishing a bound on $P(E=A)$. Informally, the difference between these two approaches is that, in the former, one is bounding the amount of information an eavesdropper can infer about agents' outcomes from the result of their chosen measurement, and, in the latter, one is bounding the probability that the eavesdropper correctly guesses agents' outcomes.

Each of the above three points will now be illustrated with concrete examples involving \emph{repeater} and \emph{star} networks. The derivations of all results in the remainder of the paper are presented in the Appendix. 


\paragraph*{Repeater networks.}Consider a \emph{repeater} network in which $n$ sources are each shared between two out of $n+1$ agents, who each perform local measurements with their devices. The crucial information about agents' inputs and outputs is captured by the DAG of Fig.~\ref{Repeater network}(a). 
The devices held by agents $A_1$ and $A_{n+1}$ have two inputs, denoted $x_1$ and $x_{n+1}$, with $x_1,x_{n+1}\in\{0,1\}$, and two possible outputs $A_1=a_1$ and $A_{n+1}=a_{n+1}$, again with $a_1,a_{n+1}\in\{0,1\}$. All remaining agents have devices with a single input and four possible outputs, denoted $A_i=a_i^0a_i^1$ with $a_i^j\in\{0,1\}$. If all $\lambda_i$ from Fig.~\ref{Repeater network}(a) are classical random variables, then an inequality bounding the classically achievable correlations is:
\begin{equation} \label{n-local inequality}
\mathcal{R}:=\sqrt{I}+\sqrt{J} \leq 1,
\end{equation}
where $I=\frac{1}{4}\sum_{x_1,x_{n+1}} \langle A_1 A_2^0 \cdots A_n^0 A_{n+1} \rangle,$
\begin{equation} \label{I and J}
\begin{aligned}
J&=\frac{1}{4}\sum_{x_1,x_{n+1}} (-1)^{x_1+x_{n+1}} \langle A_1 A_2^1 \cdots A_n^1 A_{n+1} \rangle, \\
\text{and }
 \langle & A_1 A_2^{x_2} \cdots A_{n+1} \rangle = \sum (-1)^{a_1+a_{n+1}+\sum_{i=2}^n a_i^{x_i}} \\
 &\qquad\quad \qquad \quad P(a_1, a_2^0a_2^1,\dots, a_{n+1}|x_1,x_{n+1}),
 \end{aligned}
\end{equation} 
where the above sum ranges over $a_1,a_{n+1},\dots, a_n^0a_n^1$. Inequality~(\ref{n-local inequality}) was derived in \cite{mukherjee2015correlations, branciard2010characterizing, branciard2012bilocal}, and is the analogue of a Bell inequality for this particular DAG. Note that is non-linear in the joint conditional probability distribution---hence the name \emph{polynomial} Bell inequality.

Now, if all sources are claimed by the eavesdropper to emit singlet states $\ket{\psi^-}$, devices held by agents $A_1$ and $A_{n+1}$ to be carrying out measurements $\left(\sigma_z+\sigma_x\right)/\sqrt{2}$ for $x_1=0=x_{n+1}$ and $\left(\sigma_z-\sigma_x\right)/\sqrt{2}$ for $x_1=1=x_{n+1}$ and all remaining devices to be carrying out Bell state measurements (BSMs), the generated correlations are \cite{mukherjee2015correlations}:
\begin{equation} \label{quantum correlation in repeater network} 
\begin{aligned}
\frac{1+ (-1)^{a_1+a_{n+1}}\frac{\left((-1)^{\sum_{i=2}^n a_i^0} + (-1)^{\sum_{i=2}^n a_i^1+x_1+x_{n+1}}\right)}{2}}{2^{2n}}.
\end{aligned}
\end{equation}
Plugging this into Eq.~(\ref{I and J}) yields $I=J=1/2$ \cite{mukherjee2015correlations}, which results in $\mathcal{R}=\sqrt{2}>1$, a violation of Eq.~(\ref{n-local inequality}).

Suppose, as depicted in Fig.~\ref{Repeater network}(b), the eavesdropper holds $n$ independent systems each correlated with one of the $n$ sources. Furthermore, we allow the systems held by the eavesdropper to be post-quantum (that is, non-signalling), as long as the form of the DAG from Fig~\ref{Repeater network}(b) is enforced. Each system can be measured by a device with input $z_i$ and output $E_i$. As the purpose of repeater networks is to allow agents $1$ and $n+1$ to communicate, can such an eavesdropper learn the outputs of agents $A_1$ and $A_{n+1}$? 
\begin{result}
A violation of inequality~(\ref{n-local inequality}) constitutes a bound on an eavesdropper's information about agents' outcomes in for the network of Fig~\ref{Repeater network}(b): let $D\left(P,Q\right):=\frac{1}{2}\sum_x |P(x)-Q(x)|$ denote the total variational distance, this bound on the information corresponds to
\begin{equation} \label{n-local eavesdropper bound} 
\begin{aligned}
D\Big(  P(E_1\cdots & E_n | A_1, A_{n+1}x_1,x_{n+1},z_1,\dots, z_n), \\
& P(E_1|z_1) \cdots P(E_n | z_n) \Big) \leq 2\left(2-\mathcal{R}\right).
\end{aligned}
\end{equation}
\end{result} 
While the above bound is quite weak \footnote{Note that this bound is only weak when considering quantum correlations. If all agents share PR boxes and all intermediate agents are allowed perform one of two $2$-outcome measurements in a given round, then the same bound as Eq.~(\ref{n-local eavesdropper bound}) can be derived. In this case, perfect security can be established, as PR boxes can achieve $\mathcal{R}=2$. \cite{branciard2012bilocal}.}, it formally relates polynomial inequality violation to the amount of information an eavesdropper can possess about agents' outcomes. See Appendix~\ref{Proof of n-local eavesdropper bound} for a proof. It will be shown in the star network section that increasing the number of measurement settings increases the amount of violation, hence the more measurement settings one has, the more stringent the bound on the eavesdropper's information. 
 
\begin{result} 
By correlating \emph{only} the $i=1\text{ and }n$ sources, an eavesdropper can simulate the correlations of Eq.~(\ref{quantum correlation in repeater network}). 
\end{result}
By introducing such correlations, an eavesdropper can learn agents' outcomes without alerting them \footnote{That is, once the agents have announced their inputs, as occurs in key distribution \cite{barrett2012unconditionally}}. Moreover, the sources only need to be classical. 
Such a model is provided in Appendix~\ref{classically simulating quantum correlations} and shown to correspond exactly to the quantum correlations  Eq.~(\ref{quantum correlation in repeater network}). As the eavesdropper manufactured the devices and holds a copy of each source, they can infer each agent's output. 


\begin{figure*}[t] 
\begin{subfigure}
\centering
(a)\includegraphics[scale=0.15]{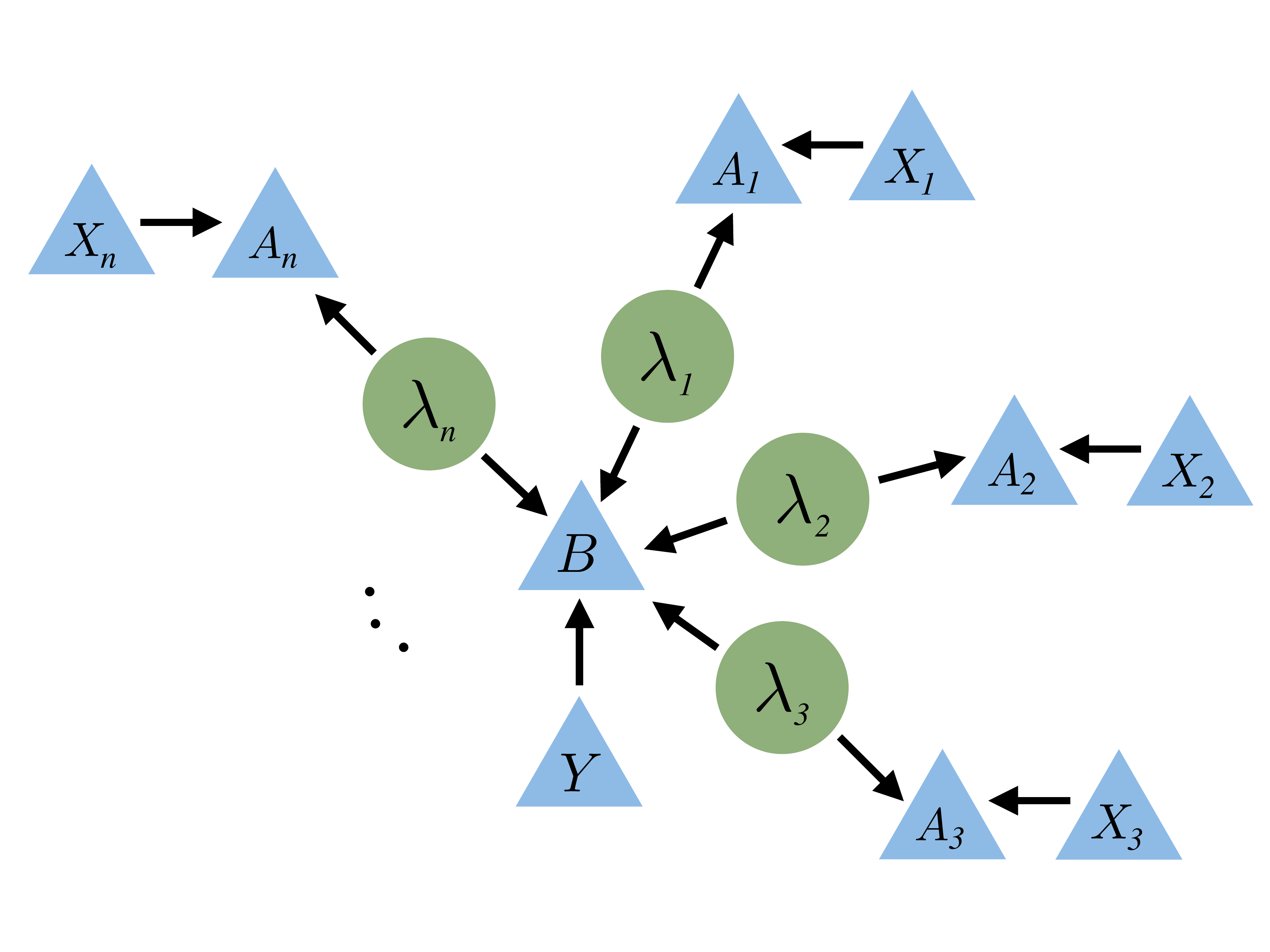} 
\end{subfigure}   
\begin{subfigure} 
\centering
(b)\includegraphics[scale=0.15]{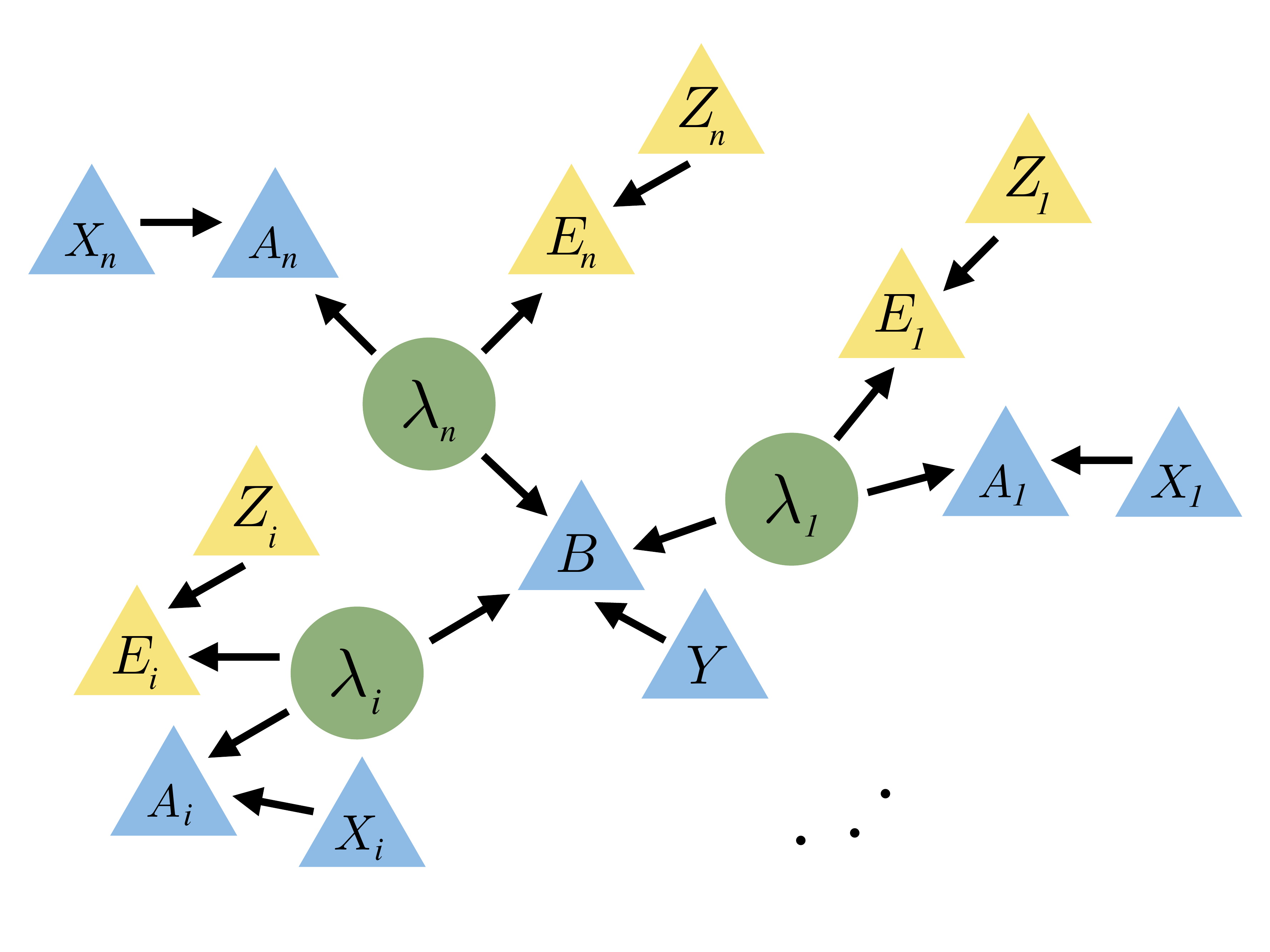}
\end{subfigure}
\begin{subfigure}
\centering
(c)\includegraphics[scale=0.15]{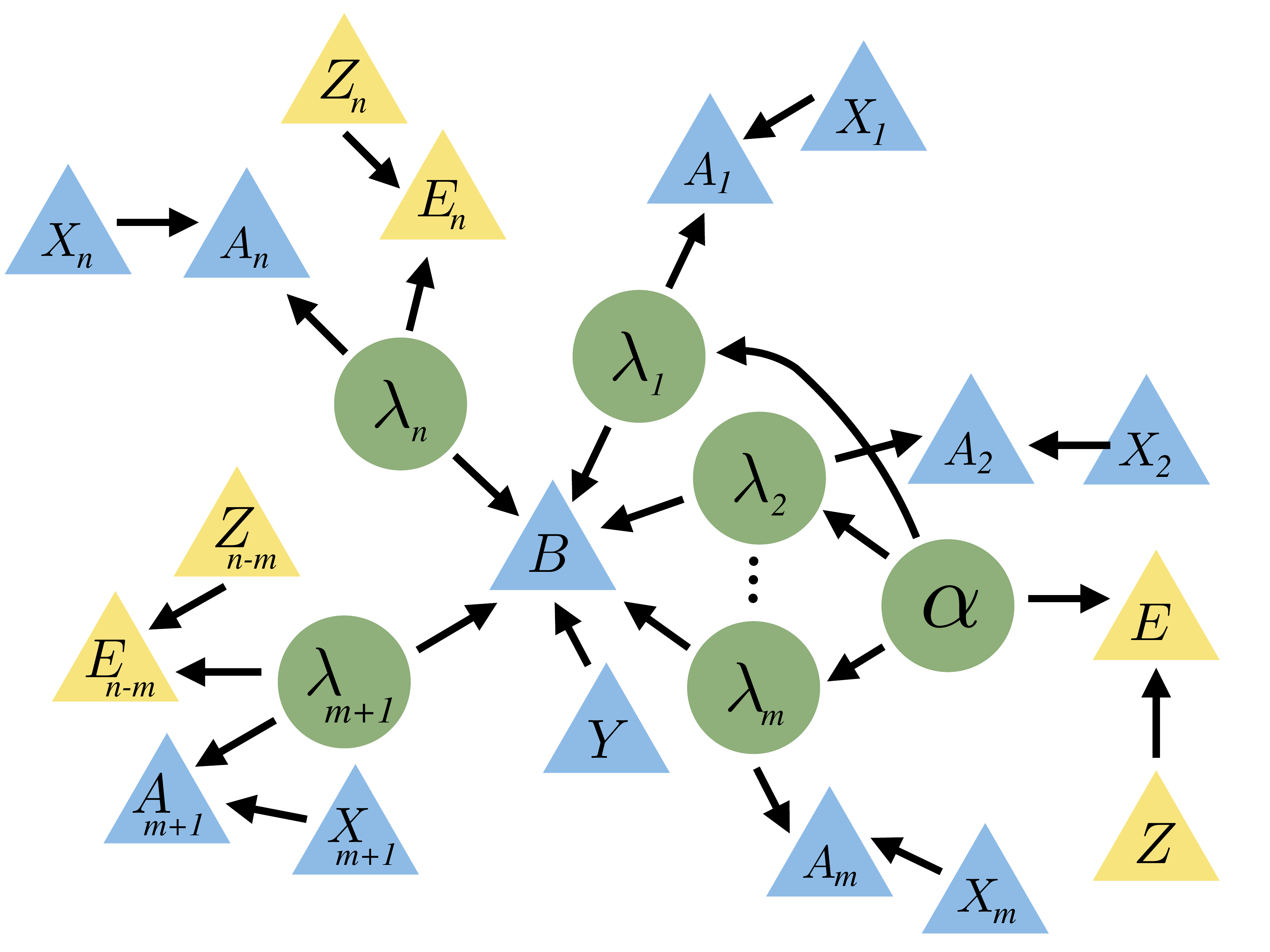}
\end{subfigure} 
\caption{
(a) Star network: $y$ \& $B$ denote input and output of central agent and $x_i$ \& $A_i$ denote the inputs and outcomes of each remaining agent. (b) Star network with eavesdropper holding system correlated with each source: $z_i$ and $E_i$ denote possible inputs and outputs of eavesdroppers' measurements. (c) Eavesdropper holding system correlated with multiple sources.
}
\label{Star network}  
\end{figure*}

There are two ways to combat this: i) having agents perform measurements that maximally violate the Clauser-Horne-Shimony-Holt (CHSH) inequality ensures their outputs are uncorrelated from an eavesdropper;  ii) take as a security assumption that the eavesdropper does not hold a system correlated with both first and last sources, as depicted in Fig.~\ref{Repeater network}(d). Even by holding a system correlated with all sources excluding the last one (or the first one), an eavesdropper cannot simulate the correlations of Eq.~(\ref{quantum correlation in repeater network}). Indeed, a variant of Eq.~(\ref{n-local eavesdropper bound}), in which the left hand side is replaced by
$D\big(  P(E_1, E_2 | A_1, \dots, z_2), P(E_1|z_1) P(E_2 | z_2) \big),$
with $E$ and $E_1$ as depicted in Fig.~\ref{Repeater network}(c), easily follows from the proof of Eq.~(\ref{n-local eavesdropper bound}) in Appendix~\ref{Proof of n-local eavesdropper bound}. 

Given the above, one might wonder why violating the CHSH inequality is not always advocated over violating inequality~(\ref{n-local inequality}). If the sources are replaced with `noisy' Bell states $\rho_i=v_i \ket{\phi^-}\bra{\phi^-} +(1-v_i)\mathbb{I}/4$, the above two cases provide an interesting trade-off between source visibility $v_i$ and security. Indeed, post-selecting intermediate BSMs results in another noisy Bell state shared between agents $A_1$ and $A_{n+1}$, but with lower visibility $V=\prod_{i=1}^n v_i$. The CHSH inequality can only be violated by this induced noisy Bell state if $V>1/\sqrt{2}$. For visibilities below this threshold no security can be established. Eq.~(\ref{n-local inequality}) however can be violated for visibilities $V>1/2$ \cite{mukherjee2015correlations, branciard2012bilocal}. Hence, noisy sources that do not ensure security in the Bell network can in principle establish some security in repeater networks.

\paragraph*{Star networks.}Consider the \emph{star} network depicted in Fig.~\ref{Star network}(a), first studied in \cite{tavakoli2014nonlocal, branciard2010characterizing, branciard2012bilocal}, consisting of $n$ independent sources, each shared between a central agent, $B$, and one of $n$ external agents, $A_i$. The devices held by each agent have $k$ possible inputs, denoted $y$ for the central agent and $x_i$ for the external agents, and two potential outputs, denoted $b$ and $a_i$. 
\begin{result}
The following inequality bounds the classically achievable correlations in Fig.~\ref{Star network}(a):
\begin{equation} \label{star network inequality}
\mathcal{S}:= \sum_{i=0}^{k-1} {|I_i|}^{1/n} \leq k-1,
\end{equation}
$\text{where, } I_i = \frac{1}{2^n}\sum_{x_1,\dots,x_n=i}^{i+1} \langle A^1_{x_1} \cdots A^n_{x_n} B_i \rangle$, for $i$ ranging from  $0 \text{ to } k-1$, with $A^i_{k}=-A^i_0$ and
$\langle A^1_{x_1} \cdots A^n_{x_n} B_y \rangle = \sum (-1)^{b+\sum_{i=2}^n a_i} P(a_1, \dots, a_{n}b|x_1,\dots, x_{n},y)$.
\end{result}
See Appendix~\ref{Proof of star network inequality} for a proof.   
  
All sources are claimed to be in the singlet state $\ket{\psi^-}$. The supplied devices are claimed to function by projecting onto the basis $ \{\cos{\frac{r\pi}{2k}}\ket{0} + \sin{\frac{r\pi}{2k}}\ket{1}, - \sin{\frac{r\pi}{2k}}\ket{0} + \cos{\frac{r\pi}{2k}}\ket{1} \}$, where $r$ is an integer equal to $x_i$ for the $i$th external agent and $y$ for central agent, outputting $0$ for the first basis element and $1$ for the second. Here, the central agent performs a separable measurement consisting of simultaneously performing the \emph{same} basis measurement on each of their joint systems, outputting the parity of individual outcomes. This implies that the operator corresponding to the the central agent's measurement factorises \footnote{This factorisation is a generalisation of the one in Eq.~($20$) of \cite{tavakoli2014nonlocal}. The derivation is the same as that of Ref.~\cite{tavakoli2014nonlocal}.} as $ B_y=B_y^1\otimes \cdots \otimes B_y^n $
resulting in $ \langle A_{x_1} \cdots A_{x_n} B_y \rangle = \langle A_x^1B_y^1 \rangle \cdots \langle A_x^nB_y^n \rangle$. Inequality~(\ref{star network inequality}) is upper bounded by $k\cos{\frac{\pi}{2k}}$ \cite{wehner2006tsirelson}. This upper bound is achieved by the above measurements \cite{braunstein1990wringing}\footnote{See the discussion in the Appendix for a proof of this}. 

In what follows the simplifying assumption that the central agents measurement device implements a separable measurement will be made. Note that, using the same approach as Section III A $3$ of \cite{tavakoli2014nonlocal}, it can been shown that all possible violations of Eq.~(\ref{star network inequality}) can be achieved using separable measurements on the center node. Hence our assumption of separable measurements is not unjustified.


Suppose, as depicted in Fig.~\ref{Star network}(b), the eavesdropper holds $n$ independent (possibly post-quantum) systems, each correlated with a single source. By using devices with inputs $z_i$ and outputs $E_i$ to measure these systems, can an eavesdropper learn the outputs of agents $A_i$? 
\begin{result}
Violating inequality~(\ref{star network inequality}) bounds an eavesdropper's information:
\begin{equation} \label{star-network eavesdropper bound} 
\begin{aligned}
D\Big( & P(E_1 \cdots  E_n | A_1,\dots, A_{n},x_1,\dots, x_{n},z_1,\dots, z_n), \\
 P(&E_1|z_1) \cdots P(E_n | z_n) \Big) \leq n\left(k-\mathcal{S}\right) \underset{k\rightarrow \infty}{\approx} \operatorname{O}\left( \frac{1}{k}\right)
\end{aligned}
\end{equation}
\end{result}  
Hence, as the number of measurement settings grows, the eavesdropper becomes increasingly uncorrelated from each agents outcome, see Appendix~\ref{Proof of star-network eavesdropper bound}. 

In the repeater network section, an eavesdropper was able to learn agents' outcomes by introducing a bit $\alpha$ which correlated the first and last sources, depicted in Fig.~\ref{Repeater network}(d). Could a similar eavesdropping attack work here? In fact, the same level of security as Eq.~(\ref{star-network correlated eavesdropper bound}) can be established against an eavesdropper who correlates $m \leq n$ sources by sharing a random variable with $q<k$ values---each taken with probability $p_l$---among the $m$ sources, as illustrated in Fig.~\ref{Star network}(c). This is formalised by demanding that an eavesdropper's information about agents' outcomes takes the following form:
\begin{equation} \label{star network correlated eavesdropper decomposition}
P_{E|A_1\cdots A_m}=\sum_{l=1}^qp_l P_{E_1^l|A_1}\cdots P_{E_m^l|A_m}.
\end{equation}

\begin{result}
Given Eq.~(\ref{star network correlated eavesdropper decomposition}), the following bound can be derived:
\begin{equation} \label{star-network correlated eavesdropper bound} 
\begin{aligned}
&D\Big( P(E E_{n-m} \cdots  E_n | A_1,x_1,\dots,z,z_{n-m}\dots, z_n), \\
 &P(E|z) \cdots P(E_n | z_n) \Big) \leq \big( n+m(q-1)\big)\left(k-\mathcal{S}\right).
\end{aligned}
\end{equation} 
\end{result}
This goes as $1/k$ for large $k$, see Appendix~\ref{Proof of star-network correlated eavesdropper bound}. If an eavesdropper introduces correlations between sources, their information of agents' outcomes can be bounded as long as the measurement settings are sufficiently large. 
  
\paragraph*{Conclusion.}
Monogamy relations using the degree of violation of a Bell inequality to bound an eavesdropper's information are central to standard device-independent information processing \cite{barrett2006maximally, colbeck2012free, aolita2012fully, augusiak2014elemental}. Thus, the results presented in this work pave the way for device-independent information processing on multi-source quantum networks. Indeed, Eq.~(\ref{quantum correlation in repeater network}) states that once intermediate outcomes are announced, agents $A_1$ and $A_{n+1}$ share a bit. Moreover, an eavesdropper's information about this bit is bounded by the degree of violation of polynomial Bell inequality~(\ref{n-local inequality}). Hence device-independent key distribution is possible in repeater networks. Moreover, security can be established using entangled sources with lower visibilities than that required for key distribution in the Bell network. Future work will focus on establishing a full security proof for device-independent key distribution on repeater networks.
 
However, for multi-source quantum networks, there are new avenues for eavesdropping attacks; by correlating sources an eavesdropper can simulate quantum correlations consistent with the original DAG. Fortunately, it was demonstrated that increasing the number of measurement settings, or ensuring the eavesdropper does not hold a system correlated with a specified subset of sources, can prevent this attack. As large-scale quantum networks---a primer for a quantum internet---are becoming possible with current technology, developing novel information processing protocols on such networks is critical. Moreover, as component networks making up future large quantum networks are likely to consist of diverse technologies, having protocols that are independent of specific technological implementations is critical.
 
Beyond generalised monogamy relations, can violation of polynomial Bell inequalities be related to advantages in other information processing tasks? Ref.'s \cite{anders2009computational, hoban2011generalized, hoban2011non} have related non-local correlations in the Bell network to quantum advantages over classical computers. This was established in the measurement-based paradigm---where adaptive measurements are performed on a single source. Relating computational advantages to violation of polynomial inequalities would be quite practical: it is more feasible to create an entangled state consisting of multiple sources of few-body entangled systems than of a single source of many-body entangled systems. 

\section*{Acknowledgements} 
CML thanks D. Browne for helpful discussions and J. J. Barry for encouragement. This work was supported by the EPSRC through the UCL Doctoral Prize Fellowship, and the Networked Quantum Information Technologies (NQIT) Hub (M013243/1). 


\appendix
\section{Brief background on DAGs} \label{DAGs}
 
The structure of each DAG encodes conditional independence relations \footnote{Here the \emph{faithfulness} condition is being assumed, see \cite{Pearl-09,wood2015lesson} for a discussion.} among the nodes. For instance, the no-signalling conditions in the Bell network, $P(A|X,Y)=P(A|X) \text{, and } P(B|X,Y)=P(B|Y)$, can be seen to directly follow \cite{wood2015lesson} from the structure of the DAG depicted in Fig.~1(a). Indeed, the specification of the DAG subsumes and generalises the standard no-signalling relations \cite{henson2014theory, fritz2012beyond, chaves2015unifying}. Indeed, in the case of the general network from Fig.~2(a), the structure of the DAG ensures $P(A_i|x_i, A_j)=P(A_i|x_i)$ for all $j\notin \{i-1, i, i+1\}$ and $P(A_i|x_i, x_j)=P(A_i|x_i)$ for all $j \neq i$. Additionally, this structure ensures $P(A_i| A_j)=P(A_i)$, for $j\notin \{i-1, i, i+1\}$. This is due to the fact that non-neighbouring agents can only be correlated through knowledge of neighbouring agents outcomes. 
Moreover, the assumption that each agent has a secure laboratory is enforced by the lack of an arrow from an eavesdropper to each agent's outcome. In short, every constraint governing how the inputs and outputs of agents and eavesdroppers are related is specified by the DAG. \color{black}

\section{Proof of Result 1} \label{Proof of n-local eavesdropper bound}
 
Consider the following conditional distribution $P(A, B, E|X,Y,Z)$, where $A,B$ are binary random variables and $X,Y$ are $k$-valued, satisfying the ``no-signalling'' conditions:
\begin{equation} \label{no-sig}
\begin{aligned}
P(A, B|X,Y,Z)&=P(A, B|X,Y) \\
P(A, E|X,Y,Z)&=P(A, E|X,Z) \\
P(B, E|X,Y,Z)&=P(B, E|Y,Z).
\end{aligned}
\end{equation}
It was shown in Ref.'s \cite{barrett2006maximally, colbeck2008hidden, colbeck2010no, barrett2012unconditionally} that
\begin{equation} \label{chained bell bound}
D\left( P(E|A, X, Z), P(E|Z) \right) \leq I_k\left( P(A, B|X,Y) \right),
\end{equation}
where $I_k$ is the chained Bell inequality \cite{braunstein1990wringing} on $k$ measurement settings, defined as:
\begin{equation} \label{chained bell inequlaity}
\begin{aligned}
I_k\left(P(A, B|X,Y)\right)&:= P(A=B|X=1, B=k)  \\
+ \sum_{|x-y|=1}& P(A \neq B | X=x, Y=y)
\end{aligned}
\end{equation}  

Consider the left hand side of inequality~(4), it will now be shown to decompose as
\begin{equation} \label{n-local bound decompoition}
\begin{aligned}
&D\Big(  P(E_1\cdots  E_n | A_1, A_{n+1}x_1,x_{n+1},z_1,\dots, z_n), \\
& P(E_1|z_1) \cdots P(E_n | z_n) \Big) \leq  D\left(P(E_1|A_1,x_1,z_1), P(E_1|z_1) \right) \\
& \qquad \qquad \qquad + D\left(P(E_n|A_n,x_n,z_n) , P(E_n|z_n)\right),
\end{aligned}
\end{equation}
where again, the systems held by the eavesdropper could be post-quantum (that is, non-signalling).
Indeed, the structure of the DAG from Fig.~2(b) implies the following conditional independence relations:
\begin{equation}
\begin{aligned}
P(E_1&\cdots  E_n | A_1, A_{n+1},x_1,x_{n+1},z_1,\dots, z_n)= \\ &P(E_1|A_1,z_1)P(E_n|A_{n+1},z_n)\prod_{i=2}^{n-1}P(E_i|z_i).
\end{aligned}
\end{equation}
Combining with the definition of $D(\cdot,\cdot)$ yields Eq.~(\ref{n-local bound decompoition}).
 
The DAG of Fig.~2(b) ensures the no-signalling relations of Eq.~(\ref{no-sig}) hold between $A_i,E_i,x_i,z_i$. The conjunction of this with Eq.'s~(\ref{chained bell bound}) and~(\ref{n-local bound decompoition}) implies
\begin{equation}\label{bounding n-local with chained bell}
\begin{aligned}
D\Big(  P(E_1\cdots & E_n | A_1, A_{n+1}x_1,x_{n+1},z_1,\dots, z_n), \\
& P(E_1|z_1) \cdots P(E_n | z_n) \Big) \leq 2 I_2 
\end{aligned}  
\end{equation}
The chained Bell inequality will now be connected to inequality~(1). First let us look at the specific case of the measurements introduced in the repeater network section. After considering this example, the general case will be proved.

To this end, consider the following mapping:
\begin{equation}\label{n-local mapping}
\begin{aligned}
P(a_1&, a_2^0a_2^1,\dots, a_{n+1}|x_1,x_{n+1}) \text{ } \longrightarrow  \\
&P(a_1, a_2,\dots,a_n, a_{n+1}|x_1,x_2, \dots, x_n, x_{n+1}) \\ 
=\sum &\delta_{a_2,a^{x_2}}\cdots \delta_{a_n,a^{x_n}} P(a_1, a_2^0a_2^1,\dots|x_1,x_{n+1}),
\end{aligned}
\end{equation}
where the sum ranges over $\{a_i^0a_i^1\}_i$. One can interpret the above mapping as follows: agents $i=2$ to $i=n$ use their devices to simulate a two choice $x_i\in\{0,1\}$, binary outcome measurement by outputting the bit $a_i^{x_i}$ from the pair output by their device $a_i^0a_i^1$. It is clear that as each agent makes their choice locally, no correlations have been introduced between agents. Hence, both $I,J$ from Eq.~(2) and $\mathcal{N}$ from Eq.~(1) are invariant under mapping~(\ref{n-local mapping}). For a more in-depth discussion on this point, see section III B from \cite{mukherjee2015correlations}.

Applying the above mapping to Eq.~(3) from the main text yields
\begin{equation} \label{seperable quantum correlation in repeater network}
\begin{aligned}
\frac{1+ (-1)^{\sum_{i=2}^n a_i}\frac{\left({\prod_{i=2}^n \delta_{x_i,0}} + (-1)^{x_1+x_{n+1}}\prod_{i=2}^n \delta_{x_i,01}\right)}{2}}{2^{n+1}}
\end{aligned}
\end{equation}
Any repeater network in which agents $2$ to $n$ have binary inputs and outputs that generates the above correlations has the same value for $\mathcal{N}$ as the one considered in the main text. Indeed, Eq.~(\ref{seperable quantum correlation in repeater network}) can be generated by agents $2$ to $n$ performing the separable measurements $A_i^y=A_{i,0}^y\otimes A_{i,1}^y\in\{\sigma_z\otimes\sigma_z, \sigma_x\otimes\sigma_x\}$, for $i=2,\dots, n$ \cite{mukherjee2015correlations}. Here, an agent measures both of their received systems in the same basis and outputs the parity of the individual measurement outcomes. Given these separable measurements for intermediate nodes, one has $\langle A_1^{x_1} A_2 \cdots A_{n+1}^{x_{n+1}} \rangle = \langle A_1^{x_2}A_{2,0}^{x_2} \rangle \langle A_{2,1}^{x_2} A_{3,0}^{x_3} \rangle \cdots \langle A_{n,1}^{x_n}A_{n+1}^{x_{n+1}} \rangle .$ Moreover, it follows that $\langle A_{i,1}^{x_i} A_{{i+1},0}^{x_{i+1}} \rangle =1$. Combining all this, it follows that
\begin{equation} \label{separable chsh bilocality}
\begin{aligned}
I&= \frac{1}{4}\left(  \langle A_1^0A_{2,0}^0 \rangle + \langle A_1^1A_{2,0}^0 \rangle \right) \left(  \langle A_{n+1}^0A_{n,1}^0 \rangle + \langle A_{n+1}^1A_{n,1}^0 \rangle \right) \\
J&= \frac{1}{4}\left(  \langle A_1^0A_{2,0}^1 \rangle - \langle A_1^1A_{2,0}^1 \rangle \right) \left(  \langle A_{n+1}^0A_{n,1}^1 \rangle - \langle A_{n+1}^1A_{n,1}^1 \rangle \right).
\end{aligned}
\end{equation}
As agents $A_1$ and $A_{n+1}$ and $A_{2,0}$ and $A_{n,1}$ respectively choose from the same set of measurements, one has
\begin{equation} \label{n-local equal to CHSH}
\begin{aligned}
\mathcal{R} = \frac{1}{2}|\langle A^0B^0 \rangle &+ \langle A^1B^0 \rangle |  \\
&+ \frac{1}{2} | \langle A^0B^1 \rangle - \langle A^1B^1 \rangle |,
\end{aligned}
\end{equation}
where  $\langle A^0B^0 \rangle := \langle A_1^0A_{2,0}^0 \rangle = \langle A_{n+1}^0A_{n,1}^0 \rangle$ formalises the statement that $A_1,A_{n+1}$ and $A_{2,0},A_{n,1}$ choose from the same set of measurements, respectively.

The above is the $k=2$ instance of the CHSH inequality
\begin{equation} \label{CHSH inequality}
\mathcal{C}_k=\sum_{i=0}^{k-1} \langle A_iB_i \rangle + \sum_{i=0}^{k-2} \langle A_{i+1}B_i \rangle -\langle A_0 B_{k-1} \rangle.
\end{equation}
Using $\langle AB \rangle = 2P(A=B)-1$, it follows that
\begin{equation} \label{relation between CHSH and chained}
I_k = k - \frac{1}{2} \mathcal{C}_k.
\end{equation}
Combing this with Eq.~(\ref{bounding n-local with chained bell}), Eq.~(4) follows.

For the general case, consider the following. As in the above example, agents $i=2$ to $i=n$ use their devices to simulate a two choice $x_i\in\{0,1\}$, binary outcome measurement by outputting the bit $a_i^{x_i}$ from the pair output by their device $a_i^0a_i^1$. This does not change the value of the polynomial Bell inequality. Moreover, as this amounts to a classical post-processing of intermediate agents outcomes, it also does not affect any potential correlations between an eavesdropper and the first and last agent in the network. 

It was shown in Theorem~$1$ of Ref.~\cite{andreoli2017maximal} that coarse-graining a two-qubit measurement with four outcomes in the manner discussed above results in a separable measurement. This is in fact always true, even for a two-qudit measurement. Indeed, as the dimension is not given a priori, without loss of generality one can always take this measurement to consist of four rank-1 projectors by appending ancillary systems. Coarse graining these as above results in operators which satisfy the conditions of Lemma~$1$ of \cite{masanes2006asymptotic}, which states that these can be reduced to a direct sum of 2-qubit operators, as is done in Eq.~($9$) of \cite{rabelo2011device} for instance. The problem has now been reduced to the case of 2-qubit measurements which were covered by \cite{andreoli2017maximal}, as discussed above.

Given intermediate separable measurements, one again has that $\langle A_1^{x_1} A_2 \cdots A_{n+1}^{x_{n+1}} \rangle = \langle A_1^{x_2}A_{2,0}^{x_2} \rangle \langle A_{2,1}^{x_2} A_{3,0}^{x_3} \rangle \cdots \langle A_{n,1}^{x_n}A_{n+1}^{x_{n+1}} \rangle .$ Following the reasoning of Eq.~(\ref{separable chsh bilocality}), one can show the following holds (see Eq.~(13) of Ref.~\cite{tavakoli2017correlations} for more details):
$$\mathcal{R} \leq \frac{1}{2} \sqrt{\mathcal{C}_k^{A_1A_{2,0}} \mathcal{C}_k^{A_{n,1}A_{n+1}}}$$
where $\mathcal{C}_k^{A_iA_j}$ is the CHSH inequality between agents $i$ and $j$. Hence, using the fact that the arithmetic mean is larger than the geometric mean, one has
$$\mathcal{R} \leq \frac{\mathcal{C}_k^{A_1A_{2,0}}}{4} + \frac{\mathcal{C}_k^{A_{n,1}A_{n+1}}}{4}.$$
Using Eq.~(\ref{relation between CHSH and chained}), one obtains 
$$2I_2 = 4 - \frac{\mathcal{C}_k^{A_1A_{2,0}}}{2} - \frac{\mathcal{C}_k^{A_{n,1}A_{n+1}}}{2} \leq 2\left( 2- \mathcal{R} \right).$$
Inputting into Eq.~(\ref{bounding n-local with chained bell}), provides the desired result.
\section{Proof of result 2: classically simulating the quantum correlations of Eq.~(3)} \label{classically simulating quantum correlations} 

It will now be demonstrated that by correlating the $i=1\text{ and }n$ sources, an eavesdropper can simulate the correlations of Eq.~(3). 
Moreover, the sources only need to emit classical variables. To achieve this, the eavesdropper sends independent and uniformly distributed bits $\{\alpha, \lambda_i\}$ to agent $A_i$, for $i=1$ and $n+1$. Given these bits and the agents' input, the agent's device outputs
$a_i=\lambda_i \oplus \alpha x_i.$
The conditional probability distribution characterising the action of the device is $P(a_i | \alpha, \lambda_i, x_i)= \frac{1}{2} \left( 1 + (-1)^{a_i+ \lambda_i + \alpha x_i} \right).$

Agent $A_i$, for $i=2\text{ and }n$, is sent independent, uniformly distributed bits $\{\alpha, \nu_i, \lambda_{i-1}, \lambda_{i}, \widetilde{\lambda}_{i}\}$, on receipt of which their device outputs
\begin{equation}
A_i = (a_i^0, a_i^1)=
\begin{cases} 
(\lambda_{i-1} \oplus \lambda_i, \nu_i), \text{ if } \alpha=0, \\
(\nu_i, \lambda_{i-1} \oplus \widetilde{\lambda}_i), \text{ if } \alpha=1.
\end{cases}
\end{equation}
Note that agents $1,2, n-1, \text{ and } n$ receive a copy of the bit $\alpha$. Hence, source $1$ and $n$ are now correlated. Recalling the transmitted bits are uniformly distributed, the conditional probability distribution characterising this device will now be derived:
$$
\begin{aligned}
&\sum_{\nu} P\left(a_i^0a_i^1 | \alpha, \nu, \lambda_{i-1},\lambda_{i}, \widetilde{\lambda}_{i}\right)P(\nu) \\
&=\sum_\nu \Big[ P(a_i^0 | \lambda_{i-1}, \lambda_i)P(a_i^1 | \nu)P(\nu)\delta_{\alpha,0} \\
& \qquad\quad\qquad+  P(a_i^1 | \lambda_{i-1}, \widetilde{\lambda}_i)P(a_i^0 | \nu)P(\nu)\delta_{\alpha,1}\Big] \\
&= \frac{1}{4}\left( 1+ (-1)^{a_i^0+\lambda_{i-1}+\lambda_i}\delta_{\alpha,0}+ (-1)^{a_i^1+\lambda_{i-1}+\widetilde{\lambda}_i}\delta_{\alpha,1}\right). 
\end{aligned}
$$

\   

Finally, all remaining agents $A_i=a_i^0a_i^1$, $i=3,\dots,n-1$, are sent uniformly distributed bits $\{\lambda_{i-1}, \widetilde{\lambda}_{i-1}, \lambda_i, \widetilde{\lambda}_i\}$. On receipt of which their devices output $a_i^0=\lambda_{i-1}\oplus\lambda_i$ and $a_i^1=\widetilde{\lambda}_{i-1}\oplus\widetilde{\lambda}_i$. The conditional probability distribution is $ \frac{1}{4}\left( 1+ (-1)^{a_i^0+\lambda_{i-1}+\lambda_i}\right) \left( 1+ (-1)^{a_i^1+\widetilde{\lambda}_{i-1}+\widetilde{\lambda}_i}\right). $ 
Combining all of these conditional probability distributions yields the following:

\begin{widetext}
$$
\begin{aligned}
P(a_1, a_2^0a_2^1,&\dots, a_n^0a_n^1,a_{n+1} | x_1, x_{n+1}) = \\
\sum_{\alpha,\nu, \lambda_1, \dots, \lambda_n} & P(a_1|\alpha,\lambda_1)P(a_2^0a_2^1|\alpha, \nu, \lambda_1, \lambda_2, \widetilde{\lambda}_2)P(a_3^0a_3^1|\lambda_{i-1}, \widetilde{\lambda}_{i-1}, \lambda_i, \widetilde{\lambda}_i) \cdots P(a_1|\alpha,\lambda_1)P(\alpha)P(\nu)P(\lambda_1)\cdots P(\lambda_n) \\
 = & \frac{1}{2^{2n}}\sum_{\alpha, \lambda_1, \dots, \lambda_n} \left( 1 + (-1)^{a_i+ \lambda_i + \alpha x_i} \right) \left( 1+ (-1)^{a_i^0+\lambda_{i-1}+\lambda_i}\delta_{\alpha,0}+ (-1)^{a_i^1+\lambda_{i-1}+\widetilde{\lambda}_i}\delta_{\alpha,1}\right)  \cdots P(\lambda_n) P(\alpha) \\
= & \frac{1}{2^{2n}}\left(1+ (-1)^{a_1+a_{n+1}}\sum_\alpha \left((-1)^{\sum_{i=2}^n a_i^0}\delta_{\alpha,0}P(\alpha) + (-1)^{\sum_{i=2}^n a_i^1+x_1+x_{n+1}}\delta_{\alpha,1}P(\alpha)\right)\right).
\end{aligned}
$$
\end{widetext}  
Performing the sum over $\alpha$ and recalling that $P(\alpha=0)=1/2=P(\alpha=1)$, results in the quantum distribution of Eq.~(3). Hence the eavesdropper can perfectly simulate quantum correlations by correlating sources thought to be independent. The last line of the above equation follows by noting that, as one multiplies out each conditional distribution, terms of the form $\sum_\lambda(-1)^\lambda$ vanish. 
\section{Proof of Result 4} \label{Proof of star network inequality}
The proof will follow the strategy of \cite[Proof of Theorem $1$]{tavakoli2014nonlocal} and \cite[Proof of Eq.~($20$)]{branciard2012bilocal}. Consider a classical model for Fig.~3(a), where all the $\lambda_i$ are random variables. Writing 
\begin{equation}
\begin{aligned}
\langle A^i_{x_i} \rangle_{\lambda_i} &= \sum_{a_i} (-1)^{a_i} P(A_i=a_i| x_i, \lambda_i) \\
\langle B_{y} \rangle_{\lambda} &= \sum_{b} (-1)^{b} P(B=b| y, \lambda),
\end{aligned}
\end{equation}
where $\lambda$ is shorthand for $\lambda_1\cdots\lambda_n$, one has 
\begin{equation}
I_i=\frac{1}{2^n}\sum_{x_1,\dots,x_n=i}^{i+1} \int \Big( \prod_{j=1}^n q_j(\lambda_j)\langle A^j_{x_j} \rangle_{\lambda_j}\Big)\langle B_{i} \rangle_{\lambda} d\lambda_j
\end{equation}
for $i=0,\dots, k-1$, where $q_j(\lambda_j)$ is the distribution over the $\lambda_i$'s. Taking the absolute value yields
\begin{equation}
|I_i| \leq \prod_{j=1}^n \Big( \frac{1}{2} \int q_j(\lambda_j) \Big| \sum_{x_j=1}^n \langle A^j_{x_j} \rangle_{\lambda_j} \Big| d\lambda_j \Big),
\end{equation}
as $\Big| \langle B_{i} \rangle_{\lambda}  \Big| \leq 1$. 

It was shown in Ref.~\cite{tavakoli2014nonlocal} that, for $c_i^k\in\mathbb{R}_+$ and $m,n\in\mathbb{N}$, the following holds:
\begin{equation}
\sum_{k=1}^m\Big( \prod_{i=1}^n c_i^k \Big)^{1/n} \leq \prod_{i=1}^{i+1} \left( c_i^1+c_i^2+\cdots + x_i^m \right)^{1/n}.
\end{equation}
Applying this result to $\mathcal{S}=\sum_{i=0}^{k-1} {|I_i|}^{1/n}$ yields
\begin{equation}
\begin{aligned}
\mathcal{S} &\leq \Big[ \prod_{j=1}^n \frac{1}{2} \int q_j(\lambda_j) \Big( \Big| \langle A^j_0 \rangle_{\lambda_j} + \langle A^j_1 \rangle_{\lambda_j} \Big|  + \Big| \langle A^j_1 \rangle_{\lambda_j} \\
&+ \langle A^j_2 \rangle_{\lambda_j} \Big|+\cdots + \Big| \langle A^j_{k-1} \rangle_{\lambda_j} - \langle A^j_0 \rangle_{\lambda_j} \Big|  \Big) d\lambda_j \Big]^{1/n}.
\end{aligned}
\end{equation}
The following upper bound holds:
\begin{equation}
\begin{aligned}
\frac{1}{2}\Big(\Big| \langle A^j_0 \rangle_{\lambda_j}& + \langle A^j_1 \rangle_{\lambda_j} \Big|  + \Big| \langle A^j_1 \rangle_{\lambda_j} + \langle A^j_2 \rangle_{\lambda_j} \Big| + \\ 
& \quad \cdots + \Big| \langle A^j_{k-1} \rangle_{\lambda_j} - \langle A^j_0 \rangle_{\lambda_j} \Big| \Big) \leq k-1.
\end{aligned}
\end{equation} 
Hence, one has
\begin{equation}
\mathcal{S} \leq \Big( \prod_{j=1}^n \int q_j(\lambda_j) \left( k-1 \right)^n d\lambda_j \Big)^{1/n}  = k-1
\end{equation}
finishing the derivation of Eq.~(5).
\section{Proof of result 5} \label{Proof of star-network eavesdropper bound}  
The structure of the DAG from Fig.~3(a) yields the following conditional independence relation:
\begin{equation} \label{conditional independence star-network}
\begin{aligned}  
P(E_1 &\cdots  E_n | A_1,\dots A_{n},x_1,\dots x_{n+1},z_1,\dots, z_n)= \\ &P(E_1|A_1, z_1)P(E_2|A_2, z_2)\cdots P(E_n|A_n, z_n).
\end{aligned}
\end{equation}
From this it follows that
$$  
\begin{aligned}
D\Big( & P(E_1 \cdots  E_n | A_1,\dots, A_{n},x_1,\dots, x_{n},z_1,\dots, z_n), \\
 P(&E_1|z_1) \cdots P(E_n | z_n) \Big) \leq \sum_i D\left(P(E_i|A_i,x_i,z_i), P(E_i|z_i) \right).
\end{aligned} 
$$ 
As stated in the main text, it is assumed that the central agents device is implementing separable measurements, hence one has $B_y=B^1_y\otimes\dots\otimes B^n_y$. Given this separable measurements, one can show that the following holds (again, see Eq.~(13) of Ref.~\cite{tavakoli2017correlations} for more details) 
\begin{equation} \label{upper bound chsh}
\mathcal{S} \leq \frac{1}{2} \left( \prod_i C_k^{A^iB^i} \right)^{\frac{1}{n}}.
\end{equation}
Following the same analysis as the end of Appendix~\ref{Proof of n-local eavesdropper bound}, the conjunction of Eq.~(\ref{upper bound chsh}) with Eq.~(\ref{chained bell bound}) 
yields Eq.~(6).

The upper bound of Eq.~(\ref{upper bound chsh}) can in fact be reached using the measurements introduced in the main paper. As all external agents choose from the same set of measurements, one has $C_k^{A^iB^i}=C_k^{A^jB^j}, \forall{i,j}$. This implies the maximal quantum value of Inequality~(5) is the maximal quantum value of $C_k^{AB}$, which has been shown by \cite{wehner2006tsirelson} to be $k\cos\left( \frac{\pi}{2k}\right).$
\section{Proof of result 6} \label{Proof of star-network correlated eavesdropper bound}
Combing Eq.~(\ref{conditional independence star-network}) with the fact that $p_l$ from Eq.~(7) satisfies $|p_l| \leq 1 \text{ } \forall i$, Eq.~(8) follows.

\bibliographystyle{ieeetr}  
\bibliography{library} 
\end{document}